\documentclass{article}
\usepackage{amssymb}

\usepackage{graphicx}
\usepackage{amsmath}

\input{tcilatex}

\begin{document}

\title{Holography\&Transplantation and All That\\
({\small The tip of an iceberg for a paradigmatic change in QFT?)}}
\author{Bert Schroer \\
CBPF, Rua Dr. Xavier Sigaud, 22290-180, Rio de Janeiro, Brazil\\
Prof. emeritus, Institut f\"{u}r Theoretische Physik, FU-Berlin\\
email: schroer@cbpf.br}
\date{June 2001}
\maketitle

\begin{abstract}
Recent developments in local quantum physics have led to revolutionary
conceptual changes in the thinking about a more intrinsic formulation and in
particular about unexpected aspects of localized degrees of freedom. This
paradigmatic change is most spectacular in a new rigorous form of
``holography'' and ``transplantation'' as generic properties in QFT beyond
the rather special geometric black hole setting in which the geometric
manifestations of these properties were first noted. This new setting is
also the natural arena for understanding the rich world of ``black hole
analogs'' (``dumb holes'' for phonons).

The mathematical basis for all this is the extremely powerful
Tomita-Takesaki modular theory in operator algebras. The rich consequences
of the impressive blend of this theory with physical localization entails
among other things the presence of ``fuzzy'' acting infinite dimensional
symmetry groups, a spacetime interpretation and derivation of the d=1+1
Zamolodchikov-Faddeev algebra (i.e. a better understanding of the
bootstrap-formfactor approach) and the noncommutative multiparticle
structure of  ``free'' anyons based on the use of Wigner representation
theory. 
\end{abstract}

\section{Introduction: important crossroads in QFT}

Genuine revolutions in theoretical physics often come in a conservative
veil, and as a result are not immediately noticed unless they lead rapidly
to new experimentally accessible predictions. In the case of Einstein's
special relativity the conservative element was the fact that not only was
the Lorentz transformation formula known before, but its incorrect
interpretation in terms of an ether-caused material contraction and dilation
effect would have created havoc with the Maxwell theory. The revolutionary
aspect of Einstein's contribution consisted in discovering a new principle
by successfully synthesizing existing principles according to their
intrinsic logic rather than cutting historical links by inventing (as
opposed to discovering) one.

The renormalized perturbation theory for quantum fields which is inexorably
linked with the names Tomonaga, Feynman Schwinger and Dyson also illustrates
this point. Apart from some modern technology (as compared e.g. to that in
Wentzel's and Heitler's prior textbooks) it is based on the old
Heisenberg-Pauli canonical quantization supplemented with the elaboration of
a remark of Kramers who reminded his colleagues of a lesson on selfmasses
(and the difference between formal Lagrangian and genuine physical
parameters in general) he learned from the Poincar\'{e} and Lorentz
treatment of classical particles within the setting of classical fields. To
be mathematically correct, Kramers remark was indispensable in removing
infinities relative to the quantization methods by which physicists in those
days entered QFT. But as the uniqueness and generic aspect of the final
finite physical (i.e. physically parametrized) results already preempted,
the intermediate infinities and the need for their removal were only caused
by the necessary repair of a slightly incorrect starting point which
warranted some intermediate ``artistic'' steps. The conceptual and
mathematical refined satisfactory treatment which removed the necessity to
deal with infinities \footnote{%
Schwinger also realized later that a careful definition of polynomial
functions of fields in terms of point split methods can avoid intermediate
infinities. Whereas these finite methods are important as a matter of
principle, they does not have the calculational efficiency of the somewhat
more artistic Feynman method.} in terms of a totally finite procedure was
later presented in the work of Lehmann, Symanzik and Zimmermann \cite{CMP}.

The revolution which is the subject of the present paper also has roots in
the history of theoretical physics, but its most recent limelight is
connected with words like holography, transmutation and scanning of data of
local quantum physics.

It is very instructive to pause and take a brief look at its historical
roots before presenting the actual setting. The before mentioned LSZ theory
combined with other developments in the early 60$^{ies}$ clarified to some
degree the relation between particles and fields and showed that the
distinction between elementary and bound/composite (even after having
specified a concrete model) is not a property which one should attribute to
particles\footnote{%
We mean particles in the conceptually precise sense of Wigner's description,
namely in terms of irreducible positive energy representations with finite
spin/helicity of the Poincar\'{e} group.} but to (generalized,
superselected) charges which can be transferred between them.

The LSZ theory did however not clarify the inverse problem (S-matrix$%
\rightarrow $QFT) nor did it answer the closely related question of which
field plays the role of what was called the ''interpolating'' field; must it
be a distinguished Lagrangian field or can it be something else? This
question was partially answered by the realization that local fields come in
equivalence classes (Borchers classes) \cite{St-Wi} and the S-matrix is to
be associated with a whole class and not with an individual member (it turns
out that there exist no physical principles which would be able to
distinguish a preferred field).

This observation was one of the theoretical pillars of algebraic QFT, the
other one was the superselection idea introduced by Wick, Wigner and
Wightman in a very special geometric context and generalized to ``charges''
by Haag and Kastler. The somewhat radical message was that the utmost
conceptual simplicity between off- and on- shell local quantum physics is
obtained by abandoning fields in favor of (nets of ) local operator
algebras. This step was similar but somewhat more radical than that from
old-fashioned coordinatized geometry to intrinsic modern coordinate-free
differential geometry\footnote{%
Whereas in differential geometry coordinate transformations between charts
are still part of the defintions, the algebraic formulation of QFT contains
no trace of field-coordinatization. }.

As was to be expected, the first success of this different viewpoint was a
purely theoretical achievement namely a profound understanding of the
internal symmetry concept associated with the action of compact groups which
arose as a generalization of Heisenberg's ``isospin''. What was at first a
strongly motivated conjecture \cite{Haag} (and an everyday experience in
Lagrangian quantization) finally turned into a deep theory. Namely it
follows from the causality and spectral principles (without using Lagrangian
quantization) in the absence of zero mass that local observables in QFTs of
spacetime dimension d$\geqslant $1+3 lead to Wigner particles and their
interpolating fields which obey the spin-statistics theorem including the
existence of a computable (from the structure of observables) compact
symmetry group acting on multiplicity indices of fields \cite{DR}. Like in
Marc Kac famous aphorism about Weyl's inverse problem : ``how to hear the
shape of a drum'', the observable ``shadow'' (which obeys the physical
causality, localization and stability properties) determines uniquely the
charged fields with their statistics (including possible braid group%
\footnote{%
In higher dimensional conformal theories there is also a conformal
decomposition theory in the timelike region leading to a braided structure
which in turn sets the spectrum of anomalous scale dimensions \cite{S}.}
statistics in low spacetime dimensions) and their multiplicity structure
including the concrete internal symmetry group (even though the observables
consisted of neutral operators which were invariant under the group action!) 
\cite{DR}. This explained in particular why in Lagrangian quantization one
was never able to find any other realization of inner symmetries and why the
low-dimensional braid group structure which leads to another symmetry
concept is out of reach for the Lagrangian formalism. The proof is
mathematically as well as conceptually very deep and a far shot removed from
the tautologies by which the Bose/Fermi alternative is ``proven'' in quantum
mechanics textbooks.

In recent times more surprising findings have been added which have no
natural description in the standard framework. But the observation which
perhaps attracts most attention to an ongoing change of paradigm in QFT is
associated with new concepts which have become known under the names
holography and transplantation. These are unexpected relation between QFTs
in different spacetime dimensions (or in the same spacetime dimension but
with different curved spacetime metric aspects), which, even if expressible
on each side in terms of pointlike fields nevertheless require a mediating
algebraic concept, which go far beyond the standard field point of view and
the Lagrangian formalism in particular. As many properties of QFT they have
been seen first through their geometric manifestations which in this case
means in the curved spacetime setting. They where so startling that their
protagonist \cite{Hooft} proposed them as a characteristic manifestation of
the elusive Quantum Gravity. In analogy with the Maldacena versus Rehren
AdS-CQFT controversy \cite{Smolin} we will show that many of these
properties are preempted in the setting of locality and spectral principles 
usual QFT.

This in my view a quite revolutionary change of paradigm in QFT (which
shares with the above mentioned discovery of renormalized perturbation and
the LSZ enrichment of the particle-field relation its strong historical ties
with previously known principles) will be the main subject of these notes.
The reason why I follow Borchers \cite{JMP} in using the adjective
``revolutionary'' can be best understood in a historical context by citing
examples for which I would not use this terminology.

Let me illustrate this by recalling an anecdotal episode in the early 60s.
Although at that time it was already known that the a one-to-one
particle-field relation was an illusion of the perturbative use of the
Lagrangian formalism, this knowledge was not yet part of the general
consciousness of particle physicist. Rather the idea that field operators
come like classical fields with an distinct intrinsic meaning was the
prevalent mode of thinking. This required to deal with rather involved
Lagrangians in which these fields (mesons, baryons or their constituents)
feature as fundamental Lagrangian fields. J. J. Sakurai \cite{Sakurai}, one
of the leading particle physicists of those years, emphasized the importance
to incorporate vectormesons and addressed but did not solve the problems of
renormalizability. At that time I was a junior collaborator of Rudolf Haag
and as such I already had acquired some familiarity with the idea of local
equivalence classes of interpolating fields associated with one particle. \
The origin of this difference to classical Lagrangian fields was that
whereas a classical particle was a separate entity and had to be imposed
``by hand'' into the classical field theory (and the prize to pay were those
infinite selfenergies observed at the beginning of last century by Poincar%
\'{e} and Lorentz), the concept of a Wigner particle was already preempted
by quantum fields as a result of the discrete irreducible component of the
positive energy representation of the Poincar\'{e} group which is
democratically attached to every individual field in the local equivalence
class of all fields carrying the same superselected charge. When Murray
Gell-Mann in 1961 was passing through the University of Illinois, there was
an interesting discussion with Haag on the partially conserved axial current
problem (PCAC) and its relation to the pion field. In the course of it
Murray Gell-Mann suddenly looked straight at us and in his matchless manner
of condensing complex ideas into just one short phrase he said ``you mean we
can shoot Sakurai?'', using Sakurai's name for the whole ``one particle --
one Lagrangian field'' mode of thinking.

Another crossroad also located at the particle-field joint, but this time
with a greater richness of physical consequences, is the seminal idea of
gauge theory which not only made the enigma of confined quarks more
respectable, but also helped to incorporate vectormesons into the family of
particles which possess renormalizable interactions. If one wants to call
this a revolution one should perhaps add the adjective ``unfinished''. The
reason for this suggestion becomes clear if one analyses the present
situation from slightly more local quantum physical (instead of a
differential geometric) viewpoint. \ After all the physical problem to
reconcile massive vector mesons with the power counting of renormalizability
was to overcome the obvious obstacle that $(m,s=1)$ physical particles have
an operator dimension 2 (and not 1 as their classical counterpart) which
makes any local coupling (which must be at least trilinear in order to serve
as an interaction) of operator dimension $\geq $5 nonrenormalizable. Such
couplings in the causal perturbation theory lead to an ever increasing
number of parameters (with the perturbative order) which renders them
practically useless\footnote{%
It has been suggested that they may serve as effective Lagrangians. Whereas
this may be true as a post factum phenomenological observation, it is
incorrect to think that there are Lagrangians which describe the situation
below a certain energy with a uniform $\varepsilon $- precision. For this
one would have to limit the local quantum physical ``phase space'' (i.e.
roughly speaking energy and localization) or name those observables for
which one can really establish such a control.}. It is a characteristic
limitation of the Lagrangian quantization approach that Lagrangian (i.e.
non-composite) fields $\psi $ with higher operator dimensions than $d_{\psi
}=1$ lead to such a situation whereas the underlying principles of QFT do
not indicate such a limitation to low spin fields. 

It is well-known that the adaptation of the classical gauge idea together
with the Higgs mechanism finally suggested a way to come to a renormalizable
theory where at the end physical vectormesons really do obtain their
physical dimension 2 (ignoring logarithmic corrections as usual in power
counting). Since there is however only one renormalizable theory (i.e. one
coupling strength) in which vectormesons interact with themselves and with
other fields (and not several competing couplings as in the classical
setting), it is somewhat misleading to talk about a gauge principle outside
the classical setting. A principle after all serves to select between a
number of possible local interactions and this is only needed in the
classical theory where the number of Lorentz invariant couplings increase
with increasing number of Lorentz indices, whereas in the quantum field
theory one usually considers the renormalizability as the overriding
principle (simply because anything ``quantum'', even if presently poorly
understood, is always considered as a potentially more fundamental than a
classical structure). Thus one obtains the \textit{gauge structure}
(uniqueness+semiclassical limit) \textit{as a result of renormalizability}
instead of the other way around. Indeed there exists a mathematical
implementation of this idea which compared with the gauge approach has the
advantage that the \textit{Higgs degrees of freedom are not put in by hand}
but the necessity to introduce additional bosonic physical degrees of \
freedom is a consequence of the causal perturbation theory. The mathematical
trick to lower the dimension of the vectormeson is to use a (BRS inspired)
cohomological representation on the level of the Wigner one particle theory
(this is only possible for nonzero mass of the incoming vectormesons) which
then guaranties that the BRS-like Fock space ghosts do not really
participate in the interaction even though they do their job as a
renormalization ``catalyzer'' \cite{S-D}. With other words the fact that the
ghosts do not interact among themselves and that they enter through a
cohomological argument facilitates the local descend to the massive physical
fields at the end which then form an operator algebra in a Hilbert space
(``unitarity'') like any other algebra (i.e.in which the catalyzer left no
noticable trace\footnote{%
Without the ``cohomological catalyzer'', the operator dimension of the
vectormeson field would increase with the perturbative order.}). The
intrinsic physical characteristics is not the Higgs mechanism but rather the
Schwinger-Swieca screening which is a consequence of the presence of the new
physical degrees of freedom (without the Higgs condensates, i.e. just
ordinary bosonic matter fields). Such theories can be distinguished from
massive QED-like models in which it is not possible to have both
renormalizable (polynomially bounded) and physical (electron) fields.

Massless theories in this approach are defined by a limiting procedure in
which the Higgs-like degrees of freedom decuple and the matter particles
become ``infraparticles'' (particles inexorably linked to their photon
clouds) or leave the physical spectrum which can be in principle studied in
perturbation theory (but this has not been done). 

It is presently not known whether for higher spins there exist more general
(cohomological or other) tricks which help to go beyond the
renormalizable/nonrenormalizable frontier set by power counting of the
physical degrees of freedom and maintain the free field dimensions
(necessarily $\geq 2)$ apart from logarithmic corrections. With other words
it is not known whether these frontiers set by the Lagrangian formalism are
really natural as far as the principles are concerned. Note that by
emphasizing the uniqueness of vectormeson couplings within the
renormalizability requirement we have not changed any of the physical
correlation functions obtained in the standard Higgs/gauge approach (apart
from the fact that the addition of new physical degrees of freedom was
required by quantum consistency and not by symmetry breaking a la Higgs of
something which never was a physical symmetry to start with \cite{Elitzur}).
The main advantage is primarily conceptual: the quantum uniqueness explains
(in the quasiclassical limit) the classical gauge structure 

Our viewpoint on vectormesons is similar to that in \cite{Scharf}, apart
from the emphasis on a ``quantum gauge principle'' (which obviously we do
not share) in the latter work. We think that our particle physics viewpoint
is more suited to highlight the incomplete aspect of the ``gauge
revolution'' i.e. the question of whether the ``cohomological catalyzer
trick'' to overcome the power counting barrier of Lagrangian quantization is
the tip of an iceberg of similar tricks for higher spins or more generally
whether the present Lagrangian frontier is the true frontier of local
quantum physics. It is hard to imagine that one is able to solve these
fundamental questions inside the standard Lagrangian framework.

This brief historical exposition of important cross roads in QFT is meant to
emphasize the fact that apart from an increase in computational
sophistication and elegance the basic formulation of using pointlike fields
for the description of interacting particles did not suffer any violent
changes through 70 years of QFT. On the other hand the algebraic framework
of local quantum physics offers not only new concepts but also demands a
different mathematical formalism. Compared with the mentioned historical
illustrations the change is indeed revolutionary especially to those for who
QFT has become synonymous with (Euclidean) functional integrals.

The ideas which have been denoted in a placative manner in the title of this
paper as ''holography, transplantation and all that'' however really do
amount to a change of paradigm or a revolution even though they have been
developed on the basis of the same physical principles as the standard text
book formulation of QFT. Their realization does not only require new
concepts but (different from the mentioned previous crossroads) also a
significant change of \ the mathematical formalism. The main novel point is
the conversion of the standard formulation into nets of operator algebras
indexed by spacetime regions or to start directly from the latter. By this
elimination of field coordinatization in terms of a more intrinsic
description (similar to the modern coordinate free way of doing differential
geometry) one gains an unexpected flexibility of reprocessing degrees of
freedom and changing their spacetime affiliation; this has no known
counterpart in the standard setting of pointlike fields. At the end one then
may, if desired, return to a pointlike description, but the field generators
for the reprocessed (holographic projection, transplantation) net have no
simple local relation to the original generating fields; here the analogy to
a coordinate transformation in differential geometry breaks down.

This new point of view is not only useful as a preparatory step towards a
better quantum understanding of Bekenstein-Hawking like black hole problems%
\footnote{%
In the curved spacetime context some of the more abstract ideas of
reprocessing operator algebras have classical geometric manifestations by
which they have been seen. Degrees of freedom conversion on causal horizons
in Minkowski spacetime (horizons of wedges or double cones) remain more
hidden behind the noncommutaive aspects of real time local quantum physics.}%
, but it also leads to a better access to more mundane problems possibly
related to laboratory physics as the description of ``free'' d=1+2 anyons
and plektons i.e. charge-carrying semiinfinitely localized operators
associated with braid group representations. Last not least it leads to a
vast generalization of the framework of Wigner symmetries: besides the
visible Poincar\'{e} or conformal symmetries there exists an infinite group
of fuzzy (non pointlike, not geometric) automorphisms of the net of operator
algebras which have other invariant states than the vacuum and are unitarily
implemented \cite{Modsym}. These new Wigner symmetries owe their existence
to the noncommutative structure and therefore cannot be seen by a Noether
type argument, but are nevertheless in complete harmony with the causality
and localization structure of the theory.

Whenever we want the reader to not think about QFT within the standard
textbook framework but to be more open-minded, we will add the adjective
algebraic (AQFT) or simply use ``local quantum physics'' (LQP).

\section{Limitations of standard QFT, transition to
field-coordinatization-free formulation}

The only structure in the Lagrangian quantization approach to which
holography has a remote connection is the light-front or p$\rightarrow
\infty $ frame quantization method \cite{holo}. For brevity let us look at
the simplest case: the d=1+1 light-ray ``quantization''. If we start from a
massive local free field $A(x)$ and restrict it to the lightray from
spacelike $x$ (say from inside the right spacelike wedge $W$) 
\begin{eqnarray}
A(x) &=&\frac{1}{\sqrt{2\pi }}\int \left( e^{-ipx}a(\theta )+h.c.\right)
d\theta ,\,p=m(ch\theta ,sh\theta ) \\
A(x_{+}) &:&=lim_{r\rightarrow 0}A(x),\,\,x=r(sh\chi ,ch\chi )  \notag
\end{eqnarray}
where the limit of lightray restriction is taken in such a way that the
inner product $px$ in the plane wave factor remains finite i.e. as $\chi =%
\hat{\chi}+lnr$ where $\hat{\chi}$ stays finite and the diverging $lnr$ term
is compensated by the vanishing $r$-factor. Of the two possibilities we
select the upper horizon limit $x_{-}\rightarrow 0$ where $x_{\pm }$ are the
lightray combinations. The resulting field $A(x_{+})$ has the usual infrared
divergence of a massless scalar field (better visible by returning to $%
p_{-}=me^{-\theta })$ which can be controlled either by restricting the test
functions for \ $A(x_{+})$ by imposing $\int f(x_{+})dx_{+}=0$ or by using
as the field $j=\partial _{+}A(x_{+})$ instead of $A(x_{+})$ for the
generation of observables. The restriction process on the lower horizon
leads to $A(x_{-})$ which has the same Fock space creation and annihilation
operators $a^{\#}$ as $A(x_{+}).$

The massless limit $m\rightarrow 0,\,\theta =\hat{\theta}+logm$ on the other
hand cannot be carried out on the level of operators in Fock space. Instead
one has to perform it on the level of correlation function ($\theta
\rightarrow \infty $ in the $a^{\#}(\theta )$ operators is meaningless) and
regain a Hilbert space and operators therein by (GNS-) reconstruction. The
result are the two chiral fields $A_{\pm }(x_{\pm })$ of chiral conformal
field theory (with the same infrared proviso as in the previous case) apart
from a doubling of degrees of freedom as compared to the previous case we
encounter the same fields.

The difference in the number of degrees of freedom is expected on classical
grounds; whereas for massless d=1+1 fields one needs both upper and lower
horizons of a wedge in order to specify the data inside the wedge, the
massive case requires the characteristic data on only one of the horizons $%
R_{u,l}$. In terms of the algebras generated by the fields the quantum
version reads 
\begin{equation}
\mathcal{A}(W)=\left\{ 
\begin{array}{c}
\mathcal{A}_{+}(R_{u})\otimes \mathcal{A}_{-}(R_{l}),\,\,m=0 \\ 
\mathcal{A}(R_{u})=\mathcal{A}(R_{l}),\,\,m\neq 0
\end{array}
\right.  \label{char}
\end{equation}
but it is very important to understand the precise mathematical meaning of
these operator relations and for this reason we will present the
mathematical concepts below.

The presence of interactions lead to some significant changes which
essentially render the lightray restriction in terms of fields ill-defined
and useless. The lightray quantization actually almost never worked for
interacting theories; in those cases where it was made to work be
decree/brute force, its connection with the original local theory became
completely obscure. The reason is two-fold. On the one hand the lightfront
fields suffer the same restrictions as those of the canonical quantization
in that the Kallen-Lehmann spectral functions must decrease so that $\int
\rho (\kappa ^{2})d\kappa ^{2}<\infty .$ This leaves only a few not very
interesting superrenormalizable $\phi _{2}^{2n}$ couplings whereas all
really interesting renormalizable interactions do not allow a lightray
approach. To be more specific, the lightfront formulation based on canonical
structures becomes as ``artistic'' as the spatial equal time canonical
formulation or the functional integral approach. It is well known that
neither canonical commutation relations nor functional integral
representations are true properties for the physical fields (after
renormalization); the only surviving property are the (bosonic, fermionic,
plektonic) spacelike commutations.

The second reason is even more serious and problematic even in the
superrenormalizable theories with decreasing spectral functions. The
lightfront approach should not change the physical content of the theory and
therefore one should be able to return to the standard formulation in which
causality and localization are manifest. Mass spectra without a localization
concept are not of much physical use; since all our interpretation goes
through locality; a direct interpretation in momentum space without
localization is an illusion. There is a good reason why despite more than 30
years writing about lightfront/$p\rightarrow \infty $ calculations one looks
in vain for a hint in this direction. The point is that in this particular
kind of problem the otherwise successful standard approach (based on field
coordinatizations i.e. short distance singular pointlike fields) has been
stretched beyond its limits of validity.

In order to maintain conceptual clarity and to keep things under
mathematical control, one must use a coordinate-free intrinsic formulation
of QFT (which often is referred to as AQFT or LQP). This will give the right
framework for making holographic mappings of degrees of freedom and in this
sense saves the intuitive content of the old lightfront approach. Before we
explain the modular inclusion concept which gives the mathematical
underpinning of the holographic reprocessing, we present a brief rundown on
the requirements of AQFT.

\begin{itemize}
\item  (i) \thinspace \thinspace \thinspace There is a map of \ compact
regions $\mathcal{O}$ in Minkowski space into von Neumann operator algebras $%
\mathcal{A(O)}$ which are subalgebras of all operators $\mathcal{B(H)}$ in
some Hilbert space $\mathcal{H}$: 
\begin{equation}
\mathcal{A}:\mathcal{O}\rightarrow \mathcal{A(O)}
\end{equation}
It is sufficient to fix the map on the Poincar\'{e} invariant family of
double cone regions ($V_{\pm }:$forward/backward lightcone) 
\begin{equation}
\mathcal{O}=\left( V_{+}+x\right) \cap \left( V_{-}+y\right) ,\,\,y-x\in
V_{+}
\end{equation}
The $C^{\ast }$-completion of this family yields the global $C^{\ast }$%
-algebra $\mathcal{A}_{quasi}:$%
\begin{equation}
\mathcal{A}_{quasi}=\bigcup_{\mathcal{O\in M}}\mathcal{A(O)}
\end{equation}
The $C^{\ast }$-algebras for noncompact regions are analogously defined by
inner approximation with double cones $\mathcal{O}.$ Since they are concrete
operator algebras in a common Hilbert space they have a natural von Neumann
closure $\mathcal{M}=\mathcal{M}^{\prime \prime }$. A closely related (but
independent) assumption 
\begin{equation}
\left\{ \bigcup_{a}\mathcal{M}(\mathcal{O}+a)\right\} ^{\prime \prime }=%
\mathcal{M}(M),\,\,M=Minkowski\,\,spacetime
\end{equation}
is called weak additivity.

\item  (ii)\thinspace \thinspace \thinspace \thinspace The family $\mathcal{A%
}$ forms a ``net'' i.e. a coherent (isotonic) family of local algebras: 
\begin{equation}
O_{1}\subset O_{2}\Longrightarrow \mathcal{A(O}_{1}\mathcal{)}\subset 
\mathcal{A(O}_{2}\mathcal{)}
\end{equation}
\end{itemize}

In case the local algebras represent observables one requires another
physically motivated coherence property namely Einstein causality or its
strengthened form called Haag duality 
\begin{eqnarray}
Einstein\,\,causality &:&\mathcal{A(O)\subset A(O}^{\prime }\mathcal{)}%
^{\prime }  \label{cau} \\
Haag\,\,duality &:&\mathcal{A(O)=A(O}^{\prime }\mathcal{)}^{\prime }
\end{eqnarray}

\begin{itemize}
\item  (iii)\thinspace \thinspace \thinspace \thinspace Covariance and
stability (positive energy condition) with respect to the Poincar\'{e} group 
$\mathcal{P}$. For observable nets: 
\begin{eqnarray*}
\alpha _{(a,\Lambda )}(\mathcal{A(O))} &=&\mathcal{A}(\Lambda \mathcal{O}+a)
\\
&=&AdU(a,\tilde{\Lambda})\mathcal{A(O)} \\
U(a,1) &=&e^{iPa},\,\,specP\in V_{+}
\end{eqnarray*}
where the unitaries represent the covering group $\mathcal{\tilde{P}}$ in $%
H. $ A particular case is that the P-spectrum contains the vacuum state $%
P\left| 0\right\rangle =0.$ We will call this net of algebras the vacuum net
and the Hilbert space $\overline{\mathcal{A}\left| 0\right\rangle }$ the
representation space of the vacuum sector. For thermal states this stability
requirement has to be changed \cite{Haag}.

\item  (iv) \ Time slice property (causal shadow property): Let $\mathcal{O}$
be the causal shadow region associated with a subregion $\mathcal{C(O)}$ of
a Cauchy surface $\mathcal{C}$ and let $U$ be a (timeslice) neighborhood of $%
\mathcal{C(O)}$ in $\mathcal{O}$, then 
\begin{equation}
\mathcal{A}(\mathcal{O})=\mathcal{A}(U)
\end{equation}

\item  (v) \ Phase space structure of LQP 
\begin{equation}
the\,map\,\Theta :\mathcal{A}(\mathcal{O})\rightarrow e^{-\beta P_{0}}%
\mathcal{A}(\mathcal{O})\Omega \text{\thinspace \thinspace }is\,\,nuclear%
\text{ \thinspace \thinspace }  \label{nuc}
\end{equation}
A more detailed account of this intricate property together with some
physical background will be given in the discussion of the ``degree of
freedom'' issue in the next two sections. .
\end{itemize}

Some comments on the physical ideas behind the requirements are in order.

In QM as formulated by von Neumann, the commutant of a collection of
Hermitian operators is a weakly closed algebra formed from all operators
which have a compatible measurement relation (physical interpretation of
commutant) with the given collection. In LQP the Einstein causality property
(ii) tells us that if the original collection generates all observables
which can be measured in a given spacetime region $\mathcal{O}$, then the
commensurable measurements are associated with observables in the spacelike
complement. There are two very important stronger versions of Einstein
causality: Haag duality and statistical independence. Haag duality is the
case of equality in (\ref{cau}) i.e. the totality of all commensurable
measurements is exhausted by the spacelike disjoint localized observables
(in QM such a characterization does not exist). One can show (if necessary
by suitably enlarging the local net within the same vacuum Hilbert space)
that Haag duality can always be achieved. It turns out that the inclusion of
the original in the Haag dualized net contains profound information on
``spontaneous symmetry breaking'' \cite{BDLR}, an issue which will not be
treated in this survey. The magnitude of violation of Haag duality in other
non-vacuum sectors is related to properties of their nontrivial
superselection charges whose mathematical description is done in terms of
endomorphisms $\rho $ \cite{Haag} of the net (the Jones index of the
inclusion $\rho (\mathcal{A})\subset \mathcal{A\,}$\ is a quantitative
measure). Neither Einstein causality nor Haag duality guaranty ``statistical
independence'' i.e. a tensor product structure between two spacelike
separated algebras analogous to the factorization for the inside/outside
region of a quantum mechanical quantization box. This kind of strengthening
of causality cannot be formulated in pure algebraic terms but needs
properties of states as well. It turns out that the nuclearity of the QFT
phase space in (v) is sufficient for the statistical independence property.
More details in a not to heavy mathematical setting can be found in \cite
{Operator}.

The positivity of energy is a specific formulation of stability adapted to
particle physics which deals with local excitations of a Poincar\'{e}
invariant vacuum. It goes back to Dirac's observation that if one does not
fill the bottom of the negative energy sea associated with the formal
energy-momentum spectrum of the Dirac equation, an external electromagnetic
interaction will create havoc. In case of thermal states it is the so-called
KMS condition which secures stability; this is the only change in the
adaptation to thermal physics \cite{Haag}.

The energy positivity leads via analytic properties of vacuum expectation
values to the cyclicity of the vacuum with respect to the action of $%
\mathcal{A}(\mathcal{O})$ i.e. $\overline{\mathcal{A}(\mathcal{O})\Omega }=H$
and for $\mathcal{O}$'s with a nontrivial causal complement the use of
causality also yields the absence of local annihilators i.e. $A\Omega
=0,\,A\in \mathcal{A(O)}\curvearrowright A=0.$ Both properties together are
known under the name of Reeh-Schlieder property \cite{St-Wi}. This property
is very different from what one is accustomed to in QM since it permits a
creation of a particle ``behind the moon'' (together with an antipartcle in
some other far remote region) by only executing local operations of
short-lived duration on the earth. Mathematically this is the starting point
for the Tomita-Takesaki modular theory which we will return to below. On the
physical side the attempts to make this exotic mathematical presence of a
dense set of state vectors by local operations physically more palatable has
led to insights into the profound role of the phase space structure (v) \cite
{Haag}.

Intuitively the connection with the formulation in terms of pointlike fields
is that the latter (smeared with $\mathcal{O}$-supported test functions in
order to obtain unbounded operators) are generators of an operator algebra $%
\mathcal{A}(\mathcal{O}),\,$\ but (as already known from the simpler case of
generators of noncompact Lie-groups) the devil lies in the details about
domains which will not concern us here since we would like to present the
new ideas with the minimum amount of technicalities. In particular the
problem whether each net of local observables fulfilling the above
requirements possesses generating pointlike fields is an open question. For
chiral conformal theories this has been shown \cite{Joerss}

Now we come to modular theory, which in a recent paper was referred to as a
revolutionizing tool \cite{JMP}. Since even in the setting of QFT
``modular'' occurs with different meanings, we will briefly define its
present use. As a side remark we mention that the more common use is that of
modular invariance in chiral conformal field theory (although this is not
its present meaning, a future connection to this causality-related
classification tool for certain families of 2-dimensional local models to
the present also locality-based use of the Tomita-Takesaki modular theory in
local quantum physics is by no means ruled out).

The Tomita-Takesaki modular theory describes a structural property of a
single operator algebra $\mathcal{A}$ in ``standard form'' i.e. under the
assumption that the Hilbert space contains a vector $\Omega $ with respect
to which the $\mathcal{A}$ acts in a cyclic and separating manner which
means that $\mathcal{A}\Omega $ is dense in $H$ and that $\mathcal{A}$
contains no annihilators of $\Omega $ ($\simeq $ to the denseness of $%
\mathcal{A}^{\prime }\Omega ,$ $\mathcal{A}^{\prime }$=commutant of $%
\mathcal{A}$). If the Hilbert space is separable, there exist always plenty
of such vectors. Operator algebras in standard form permit the definition of
the involutive antilinear (generally unbounded) Tomita $S$-operator which
without loss of generality can be assumed to be closed

\begin{eqnarray}
&&SA\Omega =A^{\ast }\Omega ,\,S=J\Delta ^{\frac{1}{2}}  \label{Tom} \\
&&S^{2}\subset \mathbf{1}  \notag
\end{eqnarray}
This operator relates the dense set $A\Omega $ to the dense set $A^{\ast
}\Omega $ for $A\in \mathcal{A}$ and gives an antiunitary $J$ and an
unbounded positive operator $\Delta ^{\frac{1}{2}}$ by polar decomposition $%
S=J\Delta ^{\frac{1}{2}}$ which have the following relation with the algebra

\begin{align}
& Ad\Delta ^{it}\mathcal{A}=\mathcal{A}  \label{modular} \\
& AdJ\mathcal{A}=\mathcal{A}^{\prime }  \notag
\end{align}

The nontrivial miraculous properties of this decomposition are the existence
of an automorphism $\sigma _{\omega }(t)=Ad\Delta ^{it}$ which propagates
operators within $\mathcal{A}$ (the first relation) and only depends on the
state $\omega $ (and not on the implementing vector $\Omega )$ and a that of
an antiunitary involution $J$ which maps $\mathcal{A}$ onto its commutant $%
\mathcal{A}^{\prime }$ (the second relation)$.$ An important thermal aspect
of the Tomita-Takesaki modular theory is the validity of the
Kubo-Martin-Schwinger (KMS) boundary condition \cite{Haag} 
\begin{equation}
\omega (\sigma _{t-i}(A)B)=\omega (B\sigma _{t}(A)),\,\,A,B\in \mathcal{A}
\label{KMS}
\end{equation}
i.e. the existence of an analytic function $F(z)\equiv \omega (\sigma
_{z}(A)B)$ holomorphic in the strip $-1<Imz<0$ and continuous on the
boundary with $F(t-i)=\omega (B\sigma _{t}(A)).$ The fact that the modular
theory applied to the wedge algebra has a geometric aspect (with $J$ equal
to the TCP operator times a spatial rotation and $\Delta ^{it}=U(\Lambda
_{W}(2\pi t))$) is not limited to the interaction-free theory \cite{Haag}.
These formulas are identical to the standard thermal KMS property of a
temperature state $\omega $ in the thermodynamic limit if one formally sets
the inverse temperature $\beta =\frac{1}{kT}$ equal to $\beta =-1.\,$This
thermal aspect (including Unruh's detector Gedankenexperiment for the
presence of radiation) is the Unruh-Hawking effect of quantum matter
enclosed behind event/causal horizons.

Our special case at hand, in which the algebras and the modular objects are
constructed functorially from the Wigner theory, suggest that the modular
structure for wedge algebras may always have a geometrical significance with
a fundamental physical interpretation in any QFT. This is indeed true, and
within the Wightman framework this was established by Bisognano and Wichmann 
\cite{Haag}.

\subsection{Connection with black hole analogs}

It cannot be emphasized enough that the thermal aspects of localization
including Hawking-like radiation are (contrary to widespread opinion) not an
exclusive attribute of curved spacetime physics or Poincar\'{e} invariance.
Rather they represent a very generic properties of quantum systems with
infinite degrees of freedom. They are related to the vacuum polarization
structure and the possibility of spatially localizing quantum matter at a
fixed time in such a way that there is a causal disjoint open region and
that none of the two regions contains annihilation operators of the ground
state $\Omega $ (a sufficient condition is a finite propagation speed in
LQP). In other words the field localization with respect to the ground state
should exhibit a unique field $-$state vector relation 
\begin{equation}
A(x)\longleftrightarrow A(x)\Omega   \label{relation}
\end{equation}
whose precise mathematical formulation is known under the name
Reeh-Schlieder theorem. In other words Hawking-like thermal behavior is
present whenever the ground state of an infinite degree of freedom problem
upon restriction to localized quantum matter behaves ``as if it would be the
vacuum in QFT''.

The phenomenon is an extremely generic one and it is somehow easier to
characterize the non-thermal exceptions. Take for example Schr\"{o}dinger QM
in the Fock space formulation. The restriction to $\psi $-operators
localized in a bounded spatial region at a fixed time does not lead to
virtual polarization effects since there is no difference between the
virtual (off-shell) particle number conservation and that of the real
(on-shell) one, so that the effect of vacuum polarization is ruled out. In
that case the vacuum in disjoint spacial regions simply factorizes, whereas
the relation \ref{relation} requires the virtual polarization property of
the vacuum. This is of course consistent with the absence of any finite
propagation speed and of causal horizons.

On the other hand phonons and many other nonrelativistic systems with many
degree of freedom systems do show the Hawking analog in the formation of
``dumb holes'' and alike traps for outgoing signals. In fact one finds a
whole zoo of black hole analog system in various different areas of physics 
\cite{Visser}. Whereas the structural presence of these thermal properties
is required by consistency of the theory, their experimentally accessibility
remains a matter of (dis)belief. Our present modular approach makes it easy
to agree with the highlighted generality of this thermal/radiation
phenomenon in the cited article, although the emphasis on the Lorentz group
and the metric structure as the reason instead of the local quantum physical
structure may be a bit self-defeating. It certainly goes against the
initially proclaimed spirit \ of that article that in order to understand
the universality of the phenomenon one should get away from differential
geometric black hole type arguments.

With the localization temperature understood in terms the KMS aspect of
modular theory, it is natural to inquire about localization entropy. Since
local operator algebras turn out to be of hyperfinite von Neumann type III$%
_{1}$ (i.e. algebras which unlike quantum mechanical algebras do not tensor
factorize), the von Neumann entropy is a priori not defined. However with
the so-called split property which follows from the above phase space
nuclearity requirement (\ref{nuc}) it is possible to find an eminent
physical substitute. The split property says that by surrounding a causal
horizon of a double cone $D$ with a ``collar'' of size $\delta $ one obtains
a canonical way to construct a quantum mechanical tensor-factorizing type I
operator algebra $\mathcal{B}_{\delta }$ which contains the given double
cone algebra and is contained in the larger double cone algebra $\hat{D}$
whose localization region is $D+\delta .$ In this way the vacuum fluctuation
at the horizon can be controlled.

The algebra $\mathcal{B}_{\delta }$ does not have a precise localization
region, its localization boundary within the collar of size $\delta $ is
totally fuzzy. Using factor $B_{\delta },$ algebra of all operators
factorizes in the manner well-known from the inside/outside quantization box
tensor products in quantum mechanics (in the 2nd quantized formulation) 
\begin{eqnarray}
B(H) &=&\mathcal{B}_{\delta }\otimes \mathcal{B}_{\delta }^{\prime } \\
H &=&H_{i}\otimes H_{o}  \notag \\
\Omega  &\neq &\Omega _{i}\otimes \Omega _{o}  \notag \\
i,o &:&inside,\text{\thinspace }outside  \notag
\end{eqnarray}
where outside here means the causal disjoint of $D+\delta .$ Different to
quantum mechanics, the global physical vacuum does not factorize into the
(analog of the) split vacua; rather it remains a highly entangled thermal
(with the same Hawking temperature) state with respect to the split
decomposition. The magnitude of the entanglement of the vacuum is measured
in terms of the entropy $S(\Omega ,\mathcal{B}_{\delta })$ of $\mathcal{B}%
_{\delta }$ relative to $\Omega .$ This entropy diverges in the limit $%
\delta \rightarrow 0$ which just expresses the fact that the entropy for the
original type III double cone algebra is ill-defined. In order to get an
intuitive feeling for the expected properties of this localization entropy $%
S(\Omega ,\mathcal{B}_{\delta })$ we look at similar but already well
understood vacuum polarization phenomena.

As first noticed by Heisenberg soon after the discovery of QFT (and later
elaborated and used by Euler, Weisskopf and many others), the partial
charge: 
\begin{equation}
Q_{V}\Omega =\int_{V}j_{0}(x)d^{3}x\Omega =\infty
\end{equation}
diverges as a result of uncontrolled vacuum particle/antiparticle
fluctuations at the boundary. In order to quantify this divergence one
should act with a more carefully defined ``partial charge'' on the vacuum
(s=dimension of space). This is an unbounded operator defined by smearing
with a test function 
\begin{eqnarray}
Q_{R,\delta } &=&\int j_{0}(x)f(x_{0})g_{\delta }(\frac{\mathbf{x}}{R})d^{s}x
\\
g_{\delta }(\mathbf{x}) &=&\left\{ 
\begin{array}{c}
1,\,\,\left| \mathbf{x}\right| <1 \\ 
0,\,\,\left| \mathbf{x}\right| >1+\delta
\end{array}
\right.  \notag \\
f(x_{0}) &\geq &0,\,suppf=\left\{ x_{0}\,|\,\left| x_{0}\right| <\varepsilon
\right\} ,\int fdx_{0}=1  \notag
\end{eqnarray}
The vectors $Q_{R,\delta }\Omega $ only converge weakly for R$\rightarrow
\infty $ on a dense domain. Their norms diverge as \cite{BDLR} 
\begin{align}
\left( Q_{R,\delta }\Omega ,Q_{R,\delta }\Omega \right) & \leq c(\delta
)R^{s-1}  \label{surface} \\
& \sim area  \notag
\end{align}
The surface character of this vacuum polarization is reflected in the area
behavior. The proportionality factor $c(\delta )$ depends on the collar size 
$\delta $ and diverges for $\delta \rightarrow 0.$ In a conformal field
theory $\delta $ and $R$ become related.

Since the localization entropy is also a vacuum polarization phenomenon, one
expects a similar behavior: an area law (at least for large double cone
radius $\delta $) with a coefficient which diverges with shrinking size of
the collar (the region of spatial change of the test function g in the
present analogy). An analogy is of course no replacement for a direct
calculation. It is typical for many properties related to modular
localization that they are easy to define and their existence is beyond
doubt, but they resist explicit calculation; perhaps because there are
presently no efficient calculational techniques.

Contrary to \cite{Visser}, I believe that an area proportionality of entropy
is inexorably related with a Hawking like thermal behavior i.e. one cannot
have one without the other. However Bekenstein's geometric-classical
derivation of the area law which gives a completely universal finite result
does not depend on a collar size and the particular kind of quantum matter.
This could find its explanation that the geometrical classical curved
spacetime does not encode the diverging leading term but only the
contribution to the area law which remains finite in the limit of vanishing
collar size. After all the quantum version of the classical equations from
which the Bekenstein law has been derived has to be renormalized and this
could effect equally all terms in that relation, not just the area term.
Although the concept and the definition of localization entropy as a measure
for the entanglement of the vacuum with respect to the splitting situation
is quite clear and unambiguous, it has up to now resisted explicit
calculation. In addition the interface between local quantum physics and
geometrodynamics i.e. the connection of the quantum localization entropy
with the classical Bekenstein entropy remains obscure. But the
well-understood connection between Hawking temperature and modular
localization leaves no doubt that there is a relation. It is encouraging for
somebody who does not believe in accidents that the same modular ideas which
are relevant for localization entropy also lead to the presence of infinite
dimensional fuzzy symmetry groups whose action on the holographic image is
that of chiral diffeomorphisms \cite{Modsym}. This is not far removed from
Carlip's use of the Virasoro algebra in his attempt to isolate the relevant
dynamical variables for the entropy on the horizon \cite{Carlip}. The
universality of the Bekenstein law fits nicely the idea that the holographic
projection onto the horizon ``kinematizes'' the degrees of freedom in that
it reprocesses the original degrees of freedom into a structureless chiral
theory with only (half)integer valued scale dimensions and that the original
dynamical richness has been transferred into the structure of the action of
automorphism (notably the one along the opposite lightray) on that
kinematical algebra.

\section{Reprocessing localization: holography\&transplantation}

In this section we will present the holography method which reformulates the
intuitive content of the lightfront quantization idea in such a way that
becomes conceptually acceptable and mathematically controllable. The
mathematical basis is a rather deep property of a special class of
inclusions of two (weakly closed) operator algebras $\mathcal{N}\subset $ $%
\mathcal{M}$ which act in the same Hilbert space. If both of them are
commutative i.e. for algebras of functions on a topological space as the
integration spaces in Euclidean functional integrals, then the projection $%
E: $ $\mathcal{M}\rightarrow $ $\mathcal{N}$ corresponds physically to the
well-known ``integrating out degrees of freedom'' (Wilson) or ``decimation''
(Kadanoff) procedure of the renormalization group formalism. If however the
algebras are noncommutative, then not every inclusion leads to such a
conditional expectation 
\begin{equation}
E(n_{1}mn_{2})=n_{1}E(m)n_{2}
\end{equation}
The precise condition which is equivalent to the existence of an $E$ has
been found by Takesaki \cite{Sunder} and simply states that $E($ $\mathcal{M}%
)$ exists iff the restriction of the modular group $\sigma _{\mathcal{M}%
}^{t} $ to $\mathcal{N}$ is the modular group of $\mathcal{N}$%
\begin{equation}
\sigma _{\mathcal{M}}^{t}|_{\mathcal{N}}=Ad\Delta _{\mathcal{M}}^{it}|_{%
\mathcal{N}}=\sigma _{\mathcal{N}}^{t}=Ad\Delta _{\mathcal{N}}^{it}|_{%
\mathcal{N}}
\end{equation}
This is a quite severe restriction which leads to the Jones subfactor theory
i.e. a framework which extends compact group symmetries. In QFT the
superselection theory of DHR leads to this situation.

The inclusions we need for the mathematical formulation of holography go one
step beyond this Takesaki situation in that the restriction leads only to a
one-sided compression

\begin{align}
& Ad\Delta _{\mathcal{\ \mathcal{M}}}^{it}\mathcal{\ \mathcal{N}}\subset 
\mathcal{\ \mathcal{N}} \\
t& \lessgtr 0,\pm halfsided\,\,modular
\end{align}
(when we simply say modular, we mean by convention $t<0$) We assume that $%
\cup _{t}Ad\Delta _{\mathcal{\ \mathcal{M}}}^{it}$\QTR{cal}{\ \QTR{cal}{N}}
is dense in $\mathcal{M}$ or equivalently that $\cap _{t}\Delta _{\mathcal{M}%
}^{it}$\QTR{cal}{\ \QTR{cal}{N}}$\mathcal{=}\mathbb{C\cdot }1$.

The above modular inclusion situation has in particular the consequence that
the two modular groups $\Delta _{\mathcal{M}}^{it}$ and $\Delta _{\mathcal{N}%
}^{it}$ generate a two-parametric group of translations and dilations in
which the translations have positive energy \cite{Wies}. Let us now look at
the relative commutant\ (see appendix of \cite{S-W3}). Let $(\mathcal{%
N\subset M},\Omega )$ be modular with nontrivial relative commutant. Then
consider the subspace generated by relative commutant $H_{red}\equiv 
\overline{(\mathcal{N}^{\prime }\cap \mathcal{M)}\Omega }\subset H.$ The
modular unitary group of $\mathcal{M}$ leaves this subspace invariant since $%
\Delta _{\mathcal{M}}^{it},t>0$ maps $\mathcal{N}^{\prime }\cap \mathcal{M}$
into itself by the inclusion being modular. Now consider the orthogonal
complement of $H_{red}$ in $H.$ This orthogonal complement is mapped into
itself by $\Delta _{\mathcal{M}}^{it}$ for positive $t$ since for $\psi $ be
in that subspace, then 
\begin{equation}
\left\langle \psi ,\Delta _{\mathcal{M}}^{it}(\mathcal{N}^{\prime }\cap 
\mathcal{M})\Omega \right\rangle =0\,\,for\,\,t>0.
\end{equation}
Analyticity in $t$ then gives the vanishing for all $t,$ i.e. invariance of $%
H_{red}.$

Due to Takesaki's theorem \cite{Sunder}, we can then restrict $\mathcal{M}$
to $H_{red}$ using a conditional expectation to this subspace defined in
terms of the projector $P$ onto $H_{red}$. Then 
\begin{align}
E(\mathcal{N}^{\prime }\cap \mathcal{M)}& \subset \mathcal{M}|_{\overline{(%
\mathcal{N}^{\prime }\cap \mathcal{M)}\Omega }}=E(\mathcal{M}) \\
E(\cdot )& =P\cdot P
\end{align}
is a modular inclusion on the subspace $H_{red}.$ $\mathcal{N}$ also
restricts to that subspace, and this restriction $E(\mathcal{N})$ is
obviously in the relative commutant of $E(\mathcal{N}^{\prime }\cap \mathcal{%
M)\subset }E(\mathcal{M)}$. Moreover using arguments as above it is easy to
see that the restriction is cyclic with respect to $\,\Omega $ on this
subspace. Therefore we arrive at a reduced modular ``standard inclusion'' 
\begin{equation}
(E(\mathcal{N)}\subset E(\mathcal{M}),\Omega )
\end{equation}
Standard modular inclusions are known to be isomorphic to chiral conformal
field theories \cite{GLW} i.e. they lead to the canonical construction of a
net $\left\{ \mathcal{A}(I)\right\} _{I\in \mathcal{K}}$ indexed by
intervals on the circle with the M\"{o}bius group PL(2,R) acting in correct
manner, including positive energy condition.

Let us now apply this to wedge algebras which are known to satisfy the
modular prerequisites. As before, we take the simplest case of a massive
d=1+1 theory 
\begin{eqnarray}
\mathcal{M} &=&\mathcal{A}(W) \\
\mathcal{N} &=&AdU(1)\mathcal{A}(W)  \notag
\end{eqnarray}
where $U(a)$ stands for the lightlike translations which slides $W$ along
the upper right lightray into itself i.e. $W_{a}\equiv W+a\subset W.$ \ The
positivity of the translational spectrum makes $\mathcal{A}(W_{a})\subset 
\mathcal{A}(W)$ a modular inclusion. The standardness of this inclusion i.e.
the triviality ($E(\mathcal{M})=\mathcal{M}$) of the above conditional
expectation is the quantum counterpart of the classical characteristic
property mentioned before (\ref{char}). The relative commutant is obviously
localized on the upper lightray. In fact it becomes a subalgebra of a
conformal net $\mathcal{A}(R)$. It is very important here to avoid equating
chiral conformal theories with zero mass particles. The lightray momenta $%
P_{\pm }$ always have a gapless spectrum even though the d=1+1 mass operator 
$\mathbb{M}^{2}=P_{+}P_{-}$ may possess a gap.

The surprising result of the existence of a chiral net on the upper horizon
is the presence of the new rotational symmetry which is directly related to
the possibility of compactifying the real line of chiral theories. Since the
chiral net consists of operator algebras which act in the same Hilbert
space, the associated unitary rotation operator acts also on the operators
of the original d=1+1 massive theory. Of course it does not belong to the
geometric symmetry operations which are exhausted by Poincar\'{e}
transformations. It is a fuzzy symmetry of the original net in the sense we
have recently introduced this concept \cite{Modsym}. On the other hand the
opposite lightray translation $U_{-}(a)=e^{iP_{-}a}$ which is a Poincar\'{e}
transformation in the original d=1+1 net becomes fuzzy on the chiral net.
Therefore diffeomorphisms of QFTs may become fuzzy in the holographic
processing and vice versa diffeomorphisms in the holographic image may
become fuzzy in the original spacetime indexing. There is absolutely no
possibility of understanding this ``scrambling up'' process in the presence
of interactions in the setting of pointlike fields, even though each side of
the holographic imaging may have a perfectly well-defined description in
terms of such fields (as generators of the respective net of algebras).

We mention in passing that within the same modular setting one can show that
each chiral theory (in our extended sense) has an infinite group of
diffeomorphisms of modular origin\footnote{%
If one defines chiral theories in the more limited context of chiral
decomposition of a massless d=1+1 theory with a traceless energy momentum
tensor these diffeomorphisms of the circle are there from the very start and
the modular theory only serves to show their modular origin.}. Higher
dimensional (not necessarily massive) theories as well as d=1+1 massive
models have a hidden infinite dimensional fuzzy symmetry group which is the
fuzzy analog of the chiral diffeomorphisms whose infinitesimal generators
form the centrally extended Virasoro algebra. In the holographic image this
fuzzy symmetry becomes a bona fide diffeomorphism group. It is an
interesting question whether the existence of the fuzzy translation $%
U_{-}(a) $ is the only distinction between the chiral nets in the sense of
this paper and the standard chiral field theory with a chiral
energy-momentum tensor.

Whereas in the free field case of the previous section the zero mass
conformal limit is (apart from multiplicity) the same as the lightray limit,
the interaction forces both chiral theory to be significantly different;
this can be made explicit within the setting of d=1+1 factorizing models 
\cite{Modsym}.

In d\TEXTsymbol{>}1+1 the presence of transversal directions to the say x-t
wedge complicates the problem because the modular inclusion cannot resolve
the transversal local net structure. It is not difficult to see that one
needs d-1 Lorentz transformation which tilt the standard wedge into d-1
different positions. The d-1 wedge algebras are reprocessed by the above
holography into d-1 copies of one chiral theory. It is not clear at this
stage of the development whether one should process these chiral copies into
the net structure of the d-1 light front in order to formulate a holographic
map onto the horizon of the d-dimensional wedge, or whether it is more
practical to use these d-1 chiral theories in the spirit of scanning the
original theory directly in terms of the relative position of d-1 chiral
theories.

An inclusion preserving map of a net in a given curved spacetime onto a net
of another spacetime in the same dimension is called a transplantation. An
interesting recent example is the transplantation of local quantum matter
from the SO(4,1) symmetric deSitter spacetime (dS) to a net on the only
SO(4)-symmetric Robertson-Walker world (RW). This transplantation is
achieved in terms of a bijective inclusion preserving map $\Xi $ of the net
of dS wedge algebras $\left\{ \mathcal{A}(W)\right\} _{W\in \mathcal{W}%
_{dS}} $ onto that of the RW wedge algebras $\left\{ \mathcal{A}(W)\right\}
_{W\in \mathcal{W}_{RW}}.$ Instead of explaining details, we refer to the
beautifully written paper \cite{Ro-Wa}. The construction also illustrates
two more concepts of modular origin: the \textbf{C}ondition of \textbf{G}%
eometric \textbf{M}odular \textbf{A}ction (CGMA) and the Modular Stability
Condition.

Compared with our previous 2-dimensional illustration of holography one
notices two interesting differences. On the one hand the transplantation of
dS to RW retains more geometric features in that the symmetry group SO(4) of
the ``Transplant'' is fully contained in SO(4,1), only the action of SO(4)%
\TEXTsymbol{\backslash}SO(4,1) is fuzzy. On the other hand the double-cone
algebras which are defined by intersecting wedges are trivial in $\mathcal{A}%
_{RW}$ as soon as the double-cone size becomes smaller than a certain
parameter \cite{Ro-Wa}. This of course means that the RW models constructed
in this way do not possess pointlike field generators. This raises the
interesting question whether there also exist physically admissable nets
which are not the transplants (or holographic images) of standard QFT. I do
not know such a model.

$\qquad $

\section{An exceptional case: the AdS-CQFT isomorphism}

There has been hardly any problem in particle physics which has attracted as
much attention as the problem if and in what way quantum matter in the 
\textbf{A}nti \textbf{d}e\textbf{S}itter spacetime and the one dimension
lower conformal field theories are related and whether this could possibly
contain clues about the meaning of quantum gravity. One reason why this
historically first example enjoyed such a widespread popularity as a test
case for new ideas about holography was that with some physical hindsight
and artistic abilities to read and interpret euclidean functional integrals
it can be seen within the standard setting of using field coordinatization.
It is not to difficult to see that univalued correlation functions of some
field on AdS spacetime upon suitable scaling adjustments (in the limit
approaching the boundary at spatial infinity) correspond to fields on
compactified Minkowski spacetime $\bar{M}.$ The latter property is
synonymous with the validity of Huygens principle and hence with the absence
of anomalous scale dimensions so that apart from different normalization
constants the correlation functions have a similar decomposition into
covariant rational functions on the complexified \ $\bar{M}$ as those for
composite free fields. In order to incorporate anomalous dimensions one has
to start with AdS objects which live on the covering of AdS. So conventional
field theoretic methods allow to understand the isomorphism \cite{Ma-Wi} in
the direction 
\begin{equation}
AdS_{d+1}\overset{rescaled}{\underset{limit}{\longrightarrow }}CQFT_{d}
\end{equation}
apart from a fine point which is related to causal propagation and which
will be mentioned later. The other direction of the arrow from the lower to
the higher dimensional theory is more subtle and cannot be done on
correlation functions or in terms of pointlike fields \cite{Rehren}. The
reason is that there is a certain amount of ``nonlocal scrambling'' of
degrees of freedom going on. \ A useful intuitive picture is obtained by
imagining the spacelike boundary of $AdS_{d+1}$ to form the wall of a d+1
dimensional cylinder with identified top-bottom (corresponding to the
periodic AdS-time) and the $AdS_{d+1}$ bulk filling its inside. For the
conformal theory on the boundary the natural causal building blocks are the
double cones which are conformally equivalent to wedges. The AdS wedges may
be viewed as the natural wedge-like prolongation of the conformal double
cones into the bulk; this purely geometric relation is of course a
reflection of the fact that the two worlds share despite their different
dimension the highest possible symmetry group. This intuitive picture gives
the correct idea about the isomorphism namely as a sur- and in-jective map
of the AdS wedge algebras and their double cone shadow algebras on the
boundary. The two theories would share the same Hilbert space and a common
algebraic structure and their only, but physically significant difference
would be the different spacetime indexing. This is indeed the content of a
rigorous mathematical theorem \cite{Rehren}.

There is another interesting lesson to be learned concerning the degrees of
freedom in this holographic reprocessing of higher to lower spacetime
dimensions. Intuitively one of course expects that a pointlike AdS theory
has too many degree of freedoms in order to be physically acceptable on the
conformal side \cite{my CMP}. This would show up in the breakdown of the
causal propagation. The algebraic formulation of causal propagation is that
the operator algebra localized in a piece of timeslice $T=\mathcal{O}%
^{(d-1)}\times \delta $ of thickness $\delta $ is equal to that of the
causal shadow $D_{T}$ (for $\mathcal{O}^{\left( d-1\right) }$ a sphere, $%
D_{T}$ would be a double cone) cast by $\mathcal{O}^{(d-1)}$%
\begin{equation}
\mathcal{A}(T)=\mathcal{A}(D_{T})
\end{equation}
This is evidently the local quantum algebraic adaptation of the classical
Cauchy propagation. Since our basic data consists of the net of double
cones, we should use the weak additivity property of AQFT and cover the time
slice $T$ by a family of small double cones. It is geometrically obvious
that each small double cone is the boundary projection of a AdS wedge which
contrary to the large wedge associated to the original double cone $D_{T}$ $%
\ $only modestly enters the bulk. So if we start from $T$ and move upward
into $D_{T}$ we expect more and more degrees of freedom entering
``sideways'' and spoil the causal propagation. Such a ``poltergeist QFT'' is
of course unacceptable as long as particle physics remains a science of
de-mystification (whether this is still a characteristic property of what
some physicist are presently doing is another matter). A nice explicit
illustration of this phenomenon is obtained by computing the conformal field
theory corresponding to a 5-dimensional zero mass\footnote{%
For massive free fields one has to use the covering of AdS, and the
resulting generalized free field have anomalous dimension and live on the
covering $\widetilde{M}$ of $\bar{M}.$} free AdS $\Phi _{0}$ field. It is a
generalized free field $\varphi $ with scale dimension $d_{\varphi }=2$%
\begin{eqnarray}
\left\langle \varphi (x)\varphi (y)\right\rangle &=&c\left[ \frac{1}{\left(
x-y\right) ^{2}}\right] ^{2} \\
\left\langle \varphi (x_{1})....\varphi (x_{n})\right\rangle &=&\left\{ 
\begin{array}{c}
\begin{array}{c}
\prod_{pairings}\left\langle \varphi (x_{i})\varphi (x_{j})\right\rangle
,\,\,n\,\,even \\ 
0,\,n\,\,odd
\end{array}
\end{array}
\right.  \notag
\end{eqnarray}
It is known that generalized free fields with certain increasing
Kallen-Lehmann spectral weights $\rho (\kappa ^{2})$ violate the timeslice
property and one checks that those homogeneous $\rho (\kappa ^{2})$ which
correspond to scale invariant theories with dimensions larger than the
canonical one all belong to this unphysical kind. This destroys the nice
dream of enlarging the extremely scarce set of 4-dimensional candidates of
conformal models by doing Lagrangian field theory on the AdS side. On the
other hand the Wigner representation theory for particles (0,s) of zero mass
is automatically (i.e. without enlargement of the irreducible representation
space of the Poincar\'{e} group) conformally invariant. These degrees of
freedom are too few in order to lead to pointlike fields on AdS \cite{my CMP}%
. Rather the pointlike conformal boundary fields stretch as (Nielsen-Olsen
like) strings\footnote{%
These ``kinematical'' localization strings do not have internal dynamical
degrees of freedom as those of string theory where the word string refers to
the dynamical spectrum and not to localization.} into the bulk i.e. the
configuration is constant along the string direction. There is no way to
escape the mathematical theorem that the isomorphism is inconsistent with an
imagined relation between two Lagrangian field theories as suggested in the
work of Maldacena Witten and others \cite{Ma-Wi} which is based on
nonrigorous arguments about converting functional integrals. As far as I
know, there has been no mathematical explanation how such an argument
concerning a relation between special models which is so close to a
structural theorem \cite{Rehren} and yet differs from it can be upheld.
Inconvenient mathematical theorems simply do not disappear by ignoring them.
.

These critical remarks about Lagrangian interpretations of relations between
models in different spacetime models can be extended to the idea of branes.
To be more specific, branes are pictured as a spacetime submanifolds with a
physical interpretation in an ambient spacetime which is also required to
have physical properties and not be just an auxiliary construct at the
service of the branes. The argument is analogous to the previous one; if one
starts with an ambient theory of pointlike fields then the causal
propagation is wrecked as a result of transversal degrees entering
continuously from the side, and if one starts on the brane side then one can
only maintain the ambient causal propagation via transversal kinematical
string-like degrees of freedom. Of course one can use the brane idea
(quasi)classically in a technical sense in order to produce lower
dimensional configurations from higher dimensional ones.

The closely related Klein-Kaluza idea for dimensional reduction is only
consistent in quasiclassical simplification of QFT which ignore the
characteristic phenomenon of vacuum polarizations. Most of the discussions
in the literature are either quasiclassical from the start or the removal of
the higher Fourier modes associated with the dimensions to be converted into
internal symmetries is done before the local quantum aspects are
investigated. If one would not have interchanged the ``small dimension
limit'' (the ``curling up'' procedure) with quantization but really have
taken this limit in the full correlations functions including the vacuum
polarization, then for most operators the latter would have diverged in an
uncontrollable way. In those few cases where authors tried to do it the
correct way, such divergent fluctuation prevented meaningful limits \cite
{Ford}. Besides this, the DR theory tells us that inner symmetry is nothing
else as geometrically encoded para-statistics \cite{Haag}, and
particle/field statistics is really a far cry away from transversal spacial
dimensions.

It is interesting to contrast the holographic property with those other
ideas of dimensional reductions. Whereas the first one needs the full power
of noncommutative local quantum aspects (to be able to apply modular
theory), the latter seem to make only sense (quasi)classically.

\section{An area of application: modular constructed anyons}

Revolutionary changes in theoretical physics should sooner or later lead to
experimentally verifiable consequences. Holography\&transplantation and the
other other new concepts in sections 2 and 3 as important for the structure
of particle physics as they may be can hardly be expected to have laboratory
manifestations in the near future, and even the nice black hole analogs from
acustics (dumb holes), hydrodynamics and other areas are presently more a
part of science fiction than of laboratory physics.

It is not necessary that the contact with the real world takes place in the
same area which highlighted the paradigmatic change. The only requirement
which an application of the new approach has to fulfill is that the
field-coordinate-free modular localization approach should be the essential
tool for its construction; with other words it should either be inaccessible
by Lagrangian quantization methods or a forced attempt to derive it with
standard methods should necessitate the use of additional unnatural recipes
and auxiliary inventions.

\subsection{weakness of the standard approach}

Such an area is the would-be theory of d=1+2 braid group statistics
fields/particles. Here we will accept the claim that braid group statistics
is the basis of some condensed matter effects (in particular the fractional
Hall effect) on face value. The difficulties one encounters with braid group
statistics in the standard setting are the following

\begin{itemize}
\item  It is not possible to maintain the plektonic (braided)
spin\&statistics connection in quantum mechanics, it rather requires the
presence of (virtual) vacuum polarization

\item  The prerequites for a euclidean functional integral representation of
correlation functions of plektonic operators are violated and hence a
definition of plektonic correlation functions on the basis of such
representations is not reliable.
\end{itemize}

These two statements need some explanatory comments.

It is well-known that spin\&statistics matters can only be meaningfully
investigated in relativistic QFT. Their use in the setting of quantum
mechanics is based on the observation that in the nonrelativistic limit one
maintains the spin\&statistics connection and obtains strict particle number
conservation (real and virtual). In this form the statement only applies to
d=1+3 spacetime dimensions where the only admissible statistics is
Bose/Fermi statistics. In d=1+2 braid group statistics is possible, but in
order to sustain it into the nonrelativistic limit, one has to abandon the
conservation of virtual particle number (no vacuum polarization) and retain
only the real particle number conservation. But quantum mechanics is
characterized by the absence of vacuum polarization and therefore the
nonrelativistic limit of a plektonic QFT will be a nonrelativistic QFT and
not a QM. The many quantum mechanical descriptions of anyons in terms of
Aharonov-Bohm potentials are misleading; wheras it is possible to deform the
spin-value by auxiliary vector potentials, it is not possible to maintain
the relativistically plektonic spin-statistics connection (as the usual
spin-statistics relation the derivation has to be done in the relativistic
setting no matter if the physical application is relativistic or
nonrelativistic) since QM does not allow for virtual vacuum polarization.
The rigorous proof consists in showing that in the relativistic setting of
``plektons'' (the abelian braid group representations are better known as
anyons) there can be no polarization-free-generators (PFG) which applied to
the vacuum generate a one particle state without admixture of
particle/antiparticle polarization clouds \cite{Mund}. Another way of
interpreting this result is to say that the mere sustention of
spin\&statistics (even without thinking about genuine interparticle
interactions) already requires the presence of these virtual clouds; in the
standard formalism one would have to blame this on some mysterious
interaction (which only exists to get the spin\&statistics going) whereas
the new formalism would automatically shift the cut between kinematics
(statistics) and dynamics (interactions). A third very provocative way of
presenting these observations is to say that quantum mechanics owes its
physical relevance to the existence of PFG Bose/Fermi fields which create
states which are \textit{free of vacuum polarization for any localization
size} (the usual free fields).

The second above listed difficulty is not independent of the first one.
Quantum field theories which lead to ``nonlocal'' fields i.e. fields which
neither commute nor anticommute for spacelike distances, are necessarily
more ``noncommutative''. The conceptual framework of Nelson-Symanzik-Guerra
for the derivation of EuclideanFeynman-Kac functional representations was
quite subtle (before it degenerated into the present ``path-integral cult'')
and limited to a small family of superrenormalizable interactions. Fermions
may be formally incorporated by first reading them back into classical
physics as Grassmann algebra valued objects and then subjecting them to an
appropriately adapted Euclidean machinery. Outside that narrow setting this
formalism is at best an artistic tool of exploration. Even if we take a very
permissive attitude and ignore the fact that in strictly renormalizable
theories (operator dimensions of interaction Lagrangians equal to spacetime
dimensions) the renormalized correlations are only Einstein causal but do
not fulfill canonical commutation relation nor euclidean functional integral
representation, there is the question whether this Euclidean setting applied
to Lagrangians containing e.g. Chern-Simons terms defines (a) a quantum
field theory at all (i.e. a theory of operators in a Hilbert space) and in
case it does (b) if the correlation functions obey the spin\&statistics
property. One can of course simply go ahead and with hindsight and ingenuity 
\cite{Suss} arrive at some quasiclassical consistency on the level of Berry
phases. But a quasiclassical consistency is no replacement for an
understanding of the spin\&statistics issue which requires a very
noncommutative structure. In fact as our above criticism of the quantum
mechanical approach based on the Aharonov-Bohm effect to the
spin\&statistics issue shows, there is all reason to mistrust quasiclassical
topological arguments in which the vacuum polarization structure does not
play an essential role. Whatever kind of quantum theory a Euclidean
functional integral containing a Chern-Simons term represents, it is not
sufficiently noncommutative in order to represent operators consistent with
braid group spin\&statistics theorem. This has led to adding
noncommutativity by hand or by taking some ideas from string theory \cite
{Gross}. The attempt to enforce this noncommutativity into condensed matter
physics by making spacetime noncommutative really seems to be very
farfetched, even if it is only for the purpose of obtaining an ``effective''
interaction. In the following I will sketch a very conservative constructive
approach which starts from Wigner one-particle theory and at the end leads
to operators which have the minimal vacuum polarization structure which is
necessary to maintain the spin\&statistics connection.

\subsection{The modular approach to ``free'' anyons}

The approach to plektons based on the new modular ideas leads to a well
defined canonical procedure which is presently in progress and will be
published in the near future. We will make no attempt here to explain all
the mathematical steps in such a constructive approach, but in order to
underline the ``revolutionary'' aspects of the modular localization approach
it is interesting to mention some of the physical concepts which are used to
construct ``free'' abelian braid group fields i.e. anyons (leaving aside the
more involved construction of general plektons). Here the word ``free''
means that we start from Wigner's one particle representation theory in
d=1+2 with an abelian little group and that in constructing the associated
multiparticle spaces and associated fields we seek that realization of the
spin\&statistics connection which contains no interactions i.e. is uniquely
determined by combining the Wigner representation theory with modular
localization theory.

If the spin is (half)integer then we would immediately go to the
(anti)symmetrized tensor Fock spaces and verify that the resulting free
fields solve the problem. In case of s$\neq $halfinteger such a procedure is
not only ill-motivated but even wrong (it would lead to generalized free
fields which violate spin\&statistics). The big surprise is now that this
innocent looking Wigner one-particle theory already preempts in an extremely
subtle way all the modular structural properties which we need in order to
construct ``free'' anyons. The irony is that this undeserved recent gift has
its origin in the very same properties which led to the ill-fated
Newton-Wigner localization (i.e. the impossibility to impose a
relativistically invariant localization concept modeled after the Born
x-space probability interpretation of wave functions) and the various
paradoxes under the name of Klein: the Wigner one-particle theory is
conceptually different from what one expects from a relativistic made
Schr\"{o}dinger quantum mechanics. This somewhat hidden feature has its
strongest outing in the existence of a pre-modular interpretation of the
Wigner theory: the correct relativistic way of localization is a spatial
version of the Tomita-Takesaki theory.

This pre-modular theory encodes relativistic localization into the position
of real Hilbert spaces within the complex Wigner representation space \cite
{Schroer}\cite{BGL}. One starts with the boost transformation associated
with a wedge and its reflection transformation along the rim of the wedge.
For the standard $x$-$t$ wedge $W_{0}$ these are the $\Lambda _{x-t}(\chi )$
Lorentz boost and the $x$-$t$ reflection $r_{x-t}:$ ($x,t)\rightarrow (-x$,$%
-t)$ which according to well-known theorems is represented antiunitarily in
the Wigner theory\footnote{%
In case of charged particles the Wigner theory should be suitably extended
by a particle/antiparticle doubling.}. One then starts from the unitary
boost group $u(\Lambda (\chi )$ and forms by the standard functional
calculus the unbounded ``analytic continuation''. In terms of modular
notation we define 
\begin{align}
\frak{s}& =\frak{j}\delta ^{\frac{1}{2}} \\
\frak{j}& =u(r)  \notag \\
\delta ^{it}& =u(\Lambda (-2\pi t))  \notag
\end{align}
where $u(\Lambda (\chi ))$ and $u(r)$ are the unitary/antiunitary
representations of these geometric transformations in the (doubled, if
particles are not selfconjugate) Wigner theory. Note that $u(r)$ is apart
from a $\pi $-rotation around the x-axis the one-particle version of the TCP
operator. On the other hand $\frak{s}$ is a very unusual object namely an
unbounded antilinear operator which on its domain is involutive $\frak{s}%
^{2}=1.$ The real subspace $H_{R}(W_{0})$%
\begin{equation}
H_{R}(W_{0})=\left\{ \psi \in H\,|\,\frak{s}\psi =\psi \right\}
\end{equation}
consists of momentum space wave functions which are boundary values of
analytic functions in the lower $i\pi -$strip of the rapidity variable $%
\theta $ and whose boundary value on one rim is the complex conjugate of
that on the other$.$ The -1 eigenvalues of $\frak{s}$ do not give rise to a
new problem since multiplication of the +1 eigenfunctions with $i$ convert
them into the -1 eigenfunctions. The real subspace $H_{R}(W_{0})$ is closed
in the complex Hilbert space topology and the complexification $%
H_{R}(W_{0})+iH_{R}(W_{0})$ gives a space which is dense in the complex
Wigner space $H$. This surprising fact (which is the Wigner one-particle
analog of the Reeh-Schlieder denseness of local field states in full quantum
field theory) has no parallel in any other area of quantum physics. It
suggests that the above mentioned unusual property of the $\frak{s}$%
-operator may be the vehicle by which geometric physical properties of
spacetime localization are encoded into the abstract domain properties of
unbounded operators. Indeed the application of the Poincar\'{e} group to the
subspace of the standard wedge generates a coherent net of subspaces $%
\left\{ H_{R}(W)\right\} _{W\in \mathcal{W}},\,\mathcal{W}$ the family of
all wedges. Some rather straightforward checks reveal that this
interpretation is consistent, namely in the present setting this
localization interpretation gives consistency with the net properties of the
spaces $H_{R}(\mathcal{O})$'s 
\begin{equation}
H_{R}(\mathcal{O})\equiv \cap _{W\supset \mathcal{O}}H_{R}(W)
\end{equation}
as well as with the conventional field theoretic construction using
pointlike fields where it agrees with localized covariant functions defined
in terms of support properties of Cauchy initial data (the compact localized
spaces $H_{R}(\mathcal{O})$ are however only nontrivial for s=(half)integer
in which case the Wigner rotation $R(\Lambda ,p)$ is free of cuts in the
modular strip). In the case of d=3+1 Wigner's ``continuous spin''
representation \cite{Weinberg} and for d=2+1 s$\neq $halfinteger these
double cone localization spaces turn out to be trivial and we will return to
this last case below.

In the halfinteger spin case the relation of Wigner subspaces and localized
subalgebras is accomplished with the help of the CCR or CAR functors which
map real subspaces $H_{R}(\mathcal{O})$ into von Neumann $\mathcal{A}(H_{R}(%
\mathcal{O}))$ subalgebras and which define a limited but rigorous meaning
of the word ``quantization'' 
\begin{equation}
J,\Delta ,S:\frak{=\Gamma (j,\delta ,s)}
\end{equation}
where the functorial map $\Gamma $ carries the functions of the Wigner
theory into the Weyl operators in Fock space (for the fermionic CAR-algebras
there is an additional modification). Whereas as previously explained the
``pre-modular'' operators denoted by small letters act on the Wigner space,
the modular operators $J,\Delta $ have an $Ad$ action on the von Neumann
algebras which are functorially related to the subspaces and which makes
them objects of the previously presented Tomita-Takesaki modular theory 
\begin{equation}
\mathcal{A}(\mathcal{O})=\left\{ 
\begin{array}{c}
alg\left\{ W(f)=e^{i(a(f)+h.a.)}|f\in H_{R}(\mathcal{O})\right\} \\ 
alg\left\{ \psi (f)\,|\,f\in H_{R}(\mathcal{O})\right\}
\end{array}
\right.
\end{equation}
where the first map of real subspaces into operator algebras is in terms of
the Weyl functor whereas the second map denoted by $\psi $ is based on the
CAR functor.

In the case of d=1+2 Wigner particles with spin $s\neq $halfinteger there
are some important changes. As in the case of s=$\frac{1}{2},$ the
spin-statistics phase is already preempted in the Wigner theory through the
appearance of a mismatch \cite{Schroer} between the spatial opposite which
is described by the symplectic complement $H_{R}(W)^{\prime }$ with the
geometric complement $H_{R}(W^{\prime })=u(R(\pi ))H_{R}(W).$ One can show
that this mismatch is described by a phase factor $t$ (the statistics phase) 
\begin{equation}
H_{R}(W^{\prime })=tH_{R}(W)^{\prime }
\end{equation}
This suggests that the Tomita S-operator of the associated full QFT is of
the form 
\begin{equation}
S=TJ\Delta ^{\frac{1}{2}}
\end{equation}
with $T$ a twist operator whose necessity was preempted by t in the Wigner
theory.

The compactly localized subspaces $H_{R}(\mathcal{O})$ turn our to be
trivial but the intersection of two (not oppositely localized) wedges which
defines a noncompact spacelike cone is nontrivial 
\begin{equation}
H_{R}(C)=H_{R}(W)\cap \left\{ u(R(\pi ))H_{R}(W)\right\} 
\end{equation}
where $R(\pi )$ a $\pi $-rotation leading to a 90 degree rotated wedge.
There is a tricky point in the calculation of intersections of wedges which
is related to the appearance of cuts for s$\neq $halfinteger in the Wigner
rotation $R(\Lambda ,p)$ which in the present d=1+2 anyon case is really a
nonlocal phase $\Phi (\Lambda ,p).$ This has been successfully treated in 
\cite{M}. In that work the spin\&statistics derivation for spacelike cone
localized fields of \cite{B-E} was adapted to the braid group situation.

The presence of the anyonic twist $t\neq \pm 1$ is an obstruction against
the existence of an operator $A(C)$ such that $A(C)\Omega \in H_{R}(\mathcal{%
O})+iH_{R}(\mathcal{O}).$ Such PFGs i.e. operators which upon application to
the vacuum create polarization cloud-free one particle states only exist for
wedges whereas any stronger localization in the presence of interactions or
genuine braid group statistics (instead of permutation group statistics)
necessarily causes polarization clouds which accompany the one-particle
state. The situation is somewhat similar to that of d=1+1 factorizing models
for which the Fouriertransforms of wedge-localized PFGs turn out to fulfill
a Zamolodchikov-Faddeev algebra \cite{Schroer}. Since the mass shell
inherits the ordering structure $p>q$ from the forward light cone, it is
possible to incorporate the anyonic phase $t$ into (Zamolodchikov-Faddeev
like) mass shell commutation relations 
\begin{eqnarray}
Z(p)Z(q) &=&tZ(q)Z(p),\,\,q>p \\
Z(p)Z^{\ast }(q) &=&2p_{0}\delta (\vec{p}-\vec{q})+tZ^{\ast }(q)Z(p),\,q>p 
\notag
\end{eqnarray}
As in the d=1+1 factorizing case the $Z^{\prime }s$ are the candidates for
the Fouriertransforms of wedge-localized PFGs and analog to that case one
also expects here that the desired spacelike cone localized free\footnote{%
Free in the sense of triviality of scattering. The constant anyonic phase
factor $t$ does not have the interpretation of a scattering matrix.} anyon
operators $A(C)$ come about by taking intersections of wedge algebras which
will determine the polarization clouds in terms of infinite power series in
the $Z$-operators. We hope to return to this interesting construction in a
different context.

\section{Concluding remarks}

Apparently QFT only attains its maximal naturalness and flexibility if one
passes from fields to nets of algebras since the radical reprocessing of
degrees of freedom in holography and transplantation does not seem to be
expressible in terms of field coordinatizations. Such a standpoint seems to
be important in the quest of definition of a localization entropy which
describes the entanglement of the vacuum state in the ``split situation''.
This may generate the desired generic local quantum physical prerequisite
for the universal nature of a quantum Bekenstein area law.

Furthermore there exist certain physical problems in which pointlike fields
are not the proper tool, e.g. the charge-carrying operators whose multiple
application to the vacuum creates state vectors with braid group statistics
necessarily have a noncompact extension (semiinfinite stringlike). Finally
there is the Holy Grail of a constructive approach in which the ultraviolet
problems are sidestepped by avoiding the use of pointlike fields. Although
the first attempt, the S-matrix bootstrap approach of the 60s failed, its
two dimensional version for factorizing S-matrices worked with the help of
special recipes. Recently it was shown that modular theory allows to justify
these recipes in terms of modular localization properties and to obtain in
particular a spacetime interpretation of the Zamolodchikov Faddeev algebra.
Since only general principles are used, there is the hope that the method
can also be extended to more general cases.

The field-coordinate free approach is based on modular Tomita-Takesaki
theory applied to nets of operator algebras. It is very surprising that
quantum field theory after 70 years of existence is capable to take such a
revolutionary turn while being totally faithful with its (in recent decades
unfortunately almost forgotten) causality and stability (spectral)
principles. But this conservative aspect is also the source of problems
which the present approach has with the prevalent Zeitgeist which seems to
follow more the desire for good entertainment than to drive forward particle
theory. For the large decades progress in particle theory has been looked
for by inviting nature to follow inventions like supersymmetry, string
theory, M-theory etc. instead of paying more conceptual attention to the
secrets which nature has hidden behind the already unravelled principles.

The situation became aggravated by the unclear experimental situation and
the increasing loss of independence of experimentalizers on what their
theory friends tell them (see the ``small extra dimensions'' search
experiments). I believe that in such a situation an approach which stays
reasonably close to the historical achievements in particle physics is the
most reasonable one. The difficulty of course is that with the accelerating
process of loss of knowledge relevant for particle theory (in some aspects
the knowledge of many especially young particle physicists has already
fallen behind the LSZ theory in which to some extend the relation between
fields and particles was already clarified), there is an increasing gap of
communication which makes reading of articles dealing with conceptual
problems of QFT difficult. Last not least progress on
conceptual-mathematical problems of local quantum physics demands a lot of
patience and strong links with history. This more contemplative spirit is
certainly not compatible with the speed which is characteristic for the
dominating areas of particle theory which, not unlike a French concord jet,
would crash if the velocity is not maintained above a certain level.

\textit{Acknowledgment}: These notes were written during a two week visit of
the Max-Planck Institut f\"{u}r Gravitationsphysik at the invitation of J%
\"{u}rgen Ehlers and Hermann Nicolai to whom I am indebted for the
invitation.

\textit{Note added}: Meanwhile I have placed some lecture notes on the
server which (on certain aspects) are more detailed \cite{Horizon}.


\begin{thebibliography}{99}
\bibitem{CMP}  A. Jaffe, G. Mack and W. Zimmermann, ``\textit{Harry Lehmann}%
'', Commun. Math Phys. \textbf{219}, (2001) 1 \ \ M. D\"{u}tsch and K.
Fredenhagen, Commun. Math Phys. \textbf{219}, (2001) 5, \ P. Breitenlohner
and D. Maison, Commun. Math Phys. \textbf{219}, (2001) 179

\bibitem{St-Wi}  R.F. Streater and A.S. Wightman, \textit{PCT, Spin and
Statistics and all That}, Benjamin 196DHR

\bibitem{Haag}  R. Haag, \textit{Local Quantum Physics}, Springer Verlag
(1992)

\bibitem{S}  B. Schroer, Phys. Lett. \textbf{B} \textbf{506}, (2001) 337

\bibitem{DR}  S. Doplicher and J. E. Roberts, Commun. Math. Phys. \textbf{131%
}, (1990) 51

\bibitem{Hooft}  G. 't Hooft, in Salam-Festschrift, A. Ali et al. eds.,
World Scientific 1993, 284. For a recent account in this fast developping
area see also S. de Haro Oll\'{e}\textit{, Quantum Gravity and the
Holographic Principle, }hep-th/0107032

\bibitem{Smolin}  M. Arnsdorf and L. Smolin, \textit{The Maldacena
Conjecture and Rehren Duality}, hep-th/0106073

\bibitem{JMP}  H. J. Borchers, J. Math. Phys. \textbf{41}, (2000) 3604

\bibitem{Sakurai}  J. J. Sakurai, Theory of strong interactions, Ann. Phys.
NY \textbf{11}, (1960) 1

\bibitem{S-D}  M. Duetsch and B. Schroer, J. Phys. A. Math. Gen. \textbf{33}%
, (2000) 4317, hep-th/9906089 and references therein.

\bibitem{Elitzur}  S. Elitzur, Phys. Rev. D \textbf{12}, (1975) 3978

\bibitem{Scharf}  G. Scharf, \textit{Quantum Gauge Theory: a True Ghost Story%
}, John Wiley \& Sons, Inc, 2001

\bibitem{holo}  B. Schroer and H.-W. Wiesbrock, Rev.Math.Phys. \textbf{12}
(2000) 461

\bibitem{Joerss}  M. J\"{o}rss, Lett. Math. Phys. \textbf{38}, (1996) 252

\bibitem{Visser}  M. Visser, \textit{Essential and inessential features of
Hawking radition}, hep-th/0106111 and literature quoted therein

\bibitem{BDLR}  D. Buchholz, S. Doplicher, R. Longo and J.H. Roberts, Rev.
Math. Phys. Special Issue (1992)

\bibitem{Operator}  B. Schroer, \textit{Lectures on Algebraic Quantum Field
Theory and Operator Algebras}, math-ph/0102018

\bibitem{Carlip}  S. Carlip, Class.Quant.Grav. 16 (1999) 3327

\bibitem{Sunder}  V. S. Sunder, \textit{An Invitation to von Neumann Algebras%
}, New York: Springer-Verlag 1987

\bibitem{Wies}  H.-W. Wiesbrock, Lett. in Math. Phys. \textbf{31}, (1994) 303

\bibitem{S-W3}  B. Schroer and H.-W. Wiesbrock, Rev.Math.Phys. \textbf{12}
(2000) 461

\bibitem{GLW}  D. Guido, R. Longo and H.-W. Wiesbrock, Commun. Math. Phys. 
\textbf{192}, (1998) \ 217

\bibitem{Modsym}  L. Fassarella and B. Schroer, \textit{The Fuzzy Analog of
Chiral Diffeomorphisms in higher dimensional Quantum Field Theories},
hep-th/0106064

\bibitem{Ro-Wa}  D. Buchholz, J. Mund, S. J. Summers, \textit{%
Transplantation of Local Nets and Geometric Modular Action on
Robertson-Walker Space-Times}, hep-th/0011237

\bibitem{Ma-Wi}  J. Maldacena, Adv. Theor. Math. Phys. \textbf{2}, (1998)
231, \ E. Witten, Adv. Theor. Math. Phys. \textbf{2}, (1998) 253, \ S. S.
Gubser, I. R. Klebanov, A. M. Poly akov, Phys.Lett. B428 (1998) 105

\bibitem{Rehren}  K.-H. Rehren, Ann, Henri Poincar\'{e} \textbf{1}, (2000)
607, \ K.-H. Rehren, \textit{Local Quantum Observables in the Anti de
Sitter-Conformal QFT Correespondence, }hep-th/0003120

\bibitem{Ford}  H. Yu and L. H. Ford, \textit{Quantum Lightcone Fluctuations
in Theories with Extra Dimensions, }gr-qc/0004063

\bibitem{my CMP}  B. Schroer, Commun. Math. Phys. \textbf{219}, (2001) 57

\bibitem{Suss}  L. Susskind, \textit{The Quantum Hall Fluid and
Non-Commutative Chern Simons Theory}, hep-th/0101029 and papers quoted
therein

\bibitem{Gross}  D. J. Gross and V. Periwal, \textit{String field theory,
non-commutative Chern-Simons theory and Lie algebra cohomology,\thinspace\ }%
hep-th/0106242

\bibitem{Mund}  J. Mund, Lett.Math.Phys. 43 (1998) 319

\bibitem{Schroer}  B. Schroer, J. Math. Phys. \textbf{41}, (2000) 3801 and
ealier papers of the author quoted therein

\bibitem{BGL}  R. Brunetti, D. Guido and R. Longo, \textit{First
quantization via BW property}, in preparation

\bibitem{Weinberg}  S. Weinberg, \textit{The Quantum Theory of Fields, I,}
Cambridge University Press 1995

\bibitem{M}  J. Mund, \textit{Localization Concept for Massive Particles
with ``Any'' Spin in d=1+2}, in preparation

\bibitem{B-E}  D. Buchholz and H. Epstein, FZKAA \textbf{17}, (1985) 329

\bibitem{Horizon}  B. Schroer, \textit{Lightlike Formalism versus
Holography\&Chiral Scanning}, hep-th/0108203
\end{thebibliography}
\end{document}